\documentclass[reprint,secnumarabic, amssymb, amsmath,aip,cha,groupedaddress,frontmatterverbose]{revtex4-1}  

\usepackage{graphicx}
\usepackage{amsmath,amssymb,amsfonts,mathtools}
\usepackage{ulem}
\usepackage{color}
\usepackage{multirow}
\usepackage{float}
\usepackage{soul}

\begin{document}

\title{Coherent State Mapping Ring Polymer Molecular Dynamics for Non-Adiabatic Quantum Propagations}
\author{Sutirtha Chowdhury}%
\author{Pengfei Huo}%
\email{pengfei.huo@rochester.edu}
\affiliation{Department of Chemistry, University of Rochester, 120 Trustee Road, Rochester, New York 14627, United States}%

\begin{abstract}
We introduce the coherent state mapping ring polymer molecular dynamics (CS-RPMD), a new method that accurately describes electronic non-adiabatic dynamics with explicit nuclear quantization. This new approach is derived by using coherent state mapping representation for the electronic degrees of freedom (DOF) and the ring-polymer path-integral representation for the nuclear DOF. CS-RPMD Hamiltonian does not contain any inter-bead coupling term in the state-dependent potential and correctly describes electronic Rabi oscillations. Classical equation of motion is used to sample initial configurations and propagate the trajectories from the CS-RPMD Hamiltonian. At the time equals to zero, the quantum Boltzmann distribution (QBD) is recovered by reweighting the sampled distribution with an additional phase factor. In a special limit that there is one bead for mapping variables and multiple beads for nuclei, CS-RPMD satisfies detailed balance and preserves an approximate QBD. Numerical tests of this method with a two-state model system show a very good agreement with exact quantum results over a broad range of electronic couplings.  
%
\end{abstract}

\maketitle
\section{Introduction}
Accurately simulating quantum dynamics effects, including non-adiabatic electronic transitions and nuclear quantum effects in large-scale condensed phase systems, is one of the central challenges in modern theoretical chemistry.\cite{althorpe2016}  Direct simulations of the exact quantum dynamics in these systems remains to be computationally demanding. It is thus ideal to develop trajectory-based approximate methods that scale linearly with respect to the nuclear degrees of freedom (DOF), while at the same time, accurately describe electronic non-adiabatic dynamics and nuclear quantum effects. 

Mixed quantum-classical (MQC) and semi-classical (SC) dynamics approaches have already been proved as promising methods that can accurately describe electronic non-adiabatic transitions. The widely used MQC methods include Ehrenfest dynamics, surface-hopping (SH) dynamics,\cite{tully1990,subotnik2016, wang2016} and mixed quantum-classical Liouville (MQCL) equation.\cite{MacKernan2008,kim2008,hsieh2012,hsieh2013,kapral2015} The commonly used SC methods include semi-classical initial-value representation (SC-IVR) path-integral methods\cite{miller2001, miller2009} and linearized path-integral dynamics.\cite{sun1998,shi2004,huo2011} All of these methods use classical trajectories to propagate the nuclear DOF, thus significantly reduce computational cost. However, the classical description of the nuclear dynamics causes inconsistencies between quantum and classical mechanics in MQC-based methods,\cite{tully2012} and cannot preserve the quantum initial distribution such as Wigner distribution used in SC-based methods.\cite{liu2011} These deficiencies can lead to problems such as the breakdown of detailed balance\cite{parandekar2005,schmidt2008} or zero-point energy leakage.\cite{muller1999,habershon2009}

Imaginary-time path-integral approaches, including centroid molecular dynamics (CMD)\cite{cao1994,jang1999} and ring polymer molecular dynamics (RPMD)\cite{habershon2013,craig2004} have been successfully developed and applied to investigate nuclear quantum effects and electronic non-adiabatic dynamics in large-scale simulations.  In particular, RPMD which resembles classical MD in an extended phase space, provides a convenient approach to compute quantum correlation functions and rate constants.\cite{habershon2013} In these methods, nuclear quantum statistics are captured with the imaginary-time path-integral formalism, leading to a ring-polymer classical isomorphism that describes quantum Boltzmann distribution (QBD) in the extended classical phase space. The classical evolution preserves the QBD captured by ring-polymer Hamiltonian due to the symplectic nature of classical dynamics, and will be free of the zero-point energy leakage problem. Despite its success in describing quantum effects in condensed phase, RPMD approach is limited to one-electron non-adiabatic dynamics\cite{menzeelev2011} or nuclear quantization,\cite{habershon2013} as well as the lack of the real-time electronic and nuclear coherence effects.\cite{menzeelev2011} 

Recent efforts have been focused on developing RPMD approaches with electronic-state representation, with a vision to accurately describes electronic dynamics and at the same time, preserve QBD.\cite{althorpe2016} Unfortunately, such methods are still missing in the current literature.  For example, Meanfield RPMD (MF-RPMD) approach\cite{hele2011,duke2016} preserves QBD; kinetically-constrained RPMD (KC-RPMD)\cite{menzeleev2014} preserves an approximated distribution that is close to QBD. However, they cannot properly describe electronic coherence because they do not contain explicit electronic state information. Mapping-variable RPMD (MV-RPMD)\cite{ananth2013} approach does employ explicit electronic state variables and preserves the exact QBD, but it cannot accurately capture Rabi oscillations in a bare two-state system.\cite{althorpe2016}  On the other hand, mapping CMD,\cite{liao2002} ring polymer surface-hopping (RPSH),\cite{shushkov2012,shakib2017} ring polymer Ehrenfest Dynamics,\cite{yoshikawa2013} and non-adiabatic mapping RPMD (NRPMD)\cite{richardson2013} are promising methods to provide explicit and accurate electronic dynamics. However, these approaches usually lack rigorous derivations and they do not preserve detailed balance in general.

In this paper, we rigorously derive a state-dependent ring-polymer Hamiltonian. Based on that, we develop a new RPMD approach, coherent state mapping RPMD (CS-RPMD). Using Meyer-Miller-Stock-Thoss\cite{meyer1979,stock1997,stock1999} representation in the coherent state basis, we introduce continuous fictitious phase space variables (mapping variables) to represent the discrete electronic states. Applying the usual path-integral technique,\cite{feynman1965,berne1986,ceperley1995,chandler1981} we derive the CS-RPMD partition function expression that provides exact QBD through the extended phase space description. Initial distributions in the CS-RPMD are sampled from the classical dynamics of the CS-RPMD Hamiltonian. By using coherent state basis for the mapping variables, CS-RPMD Hamiltonian does not contain any inter-bead coupling terms in the state-dependent mapping potential, leading to an accurate description of the electronic dynamics and correct Rabi oscillations. In the adiabatic limit and state-independent limit, CS-RPMD reduces back to the regular RPMD, just like any state-dependent RPMD approach,\cite{menzeleev2014,ananth2013,duke2016} and thus rigorously preserves QBD under this limit. In a special case that there is only one bead for the mapping variables but still multiple beads for the nuclear DOF, we can rigorously prove that CS-RPMD satisfies detailed balance and preserves an approximate QBD.  While the NRPMD approach assumes a Hamiltonian that closely resembles CS-RPMD Hamiltonian,\cite{richardson2013} the current work demonstrates a rigorous way to derive this Hamiltonian with a partition function that provides exact QBD, providing a solid theoretical foundation.

\section{Theory}
We start with expressing the total Hamiltonian operator of the system as follows
\begin{equation}\label{eqn:tot-H}
\hat{H}=\hat{T}+\hat{V}_{0}+\hat{H}_{e} ={\hat{\bf P}^{2}\over{2M}}+V_{0}({\bf \hat{R}})+\sum_{n,m=1}^{L}V_{nm}({\bf \hat{R}})|n\rangle \langle m|,
\end{equation}
where $\hat{T}$ is the nuclear kinetic energy operator, $\hat{\bf P}$ is the nuclear momentum operator, $M$ is the nuclear mass. $V_{0}({\bf \hat{R}})$ is the state-independent potential operator, and $\hat{H}_\mathrm{e}=\sum_{nm}V_{nm}({\bf \hat{R}})|n\rangle \langle m|$ is the state-dependent potential operator (electronic part of the Hamiltonian) with $L$ total diabatic electronic states. 

To derive the CS-RPMD Hamiltonian, we start from the canonical partition function defined as $\mathcal{Z}=\mathrm{Tr_{en}}[e^{-\beta\hat H}]$, where $\mathrm{Tr_{en}=Tr_{e}Tr_{n}}$ represents the trace over both electronic and nuclear DOFs, $\beta=1/k_\mathrm{B}T$ is the reciprocal temperature, and $\hat{H}$ is the total Hamiltonian operator defined in Eqn.~\ref{eqn:tot-H}. The partition function can be exactly evaluated as $\mathcal{Z}=\mathrm{Tr_{en}}\prod_{\alpha = 1}^{N}[e^{-\beta_{N}\hat H}]$, with $\alpha$ as the imaginary-time (bead) index, and a higher effective temperature defined as $\beta_{N}=\beta/N$. Further splitting the Boltzmann operator by trotter expansion under the infinite bead limit $N\rightarrow\infty$ gives $\mathcal{Z}=\lim_{N\to\infty} \mathrm{Tr_{en}} \prod_{\alpha = 1}^{N}[e^{-\beta_{N}(\hat{T}+\hat{V}_{0})}e^{-\beta_{N}\hat{H}_{e}}]$. Inserting $N$ copies of the resolution of identity $I_{\bf R} = \int {\bf dR_{\alpha}} |\bf {R_{\alpha}}\rangle\langle {\bf R_{\alpha}}|$ and $I_{\bf P}= \int {\bf dP_{\alpha}} |\bf {P_{\alpha}}\rangle\langle {\bf P_{\alpha}}|$, and explicitly performing the trace over the nuclear DOF based on the standard path-integral technique, \cite{feynman1965,berne1986,ceperley1995,chandler1981} we have 
\begin{equation}\label{eqn:part-gen}
\mathcal{Z}  = \lim_{N\to\infty} \int d \{{\bf P_\alpha}\} d\{{\bf R_\alpha}\} e^{-\beta_{N}H_\mathrm{rp}} \text{Tr}_\mathrm{e}\prod_{\alpha = 1}^{N} [e^{-\beta_{N}\hat {H}_{e}({\bf R_{\alpha}})}],
\end{equation}
with $\int d \{{\bf P_\alpha}\} d\{{\bf R_\alpha}\}=\prod_{\alpha = 1}^{N} \int d {\bf P_\alpha} d{\bf R_\alpha}$.
Here, the ring-polymer Hamiltonian $H_\mathrm{rp}$ is expressed as follows
\begin{equation}\label{eqn:hrp}
H_\mathrm{rp}=\sum_{\alpha=1}^{N}{{{\bf P_{\alpha}}^{2}}\over{2M}}+V_{0}({\bf R_{\alpha}})+{M\over{2\beta^{2}_{N}\hbar^{2}}}({\bf R_{\alpha}-R_{\alpha-1}})^{2},
\end{equation}
and the state-independent potential operator $\hat {H}_{e}({\bf R_{\alpha}})=\sum_{n,m}V_{nm}({\bf R_\alpha})|n\rangle\langle m|$ parametrically depends upon $\alpha_\mathrm{th}$ bead's nuclear position ${\bf R_{\alpha}}$.

The above partition function is a common expression and the starting point for all state-dependent RPMD approaches\cite{menzeleev2014,ananth2013,richardson2013,hele2011,duke2016} and path-integral Monte-Carlo (PIMC) methods.\cite{alexander2001,schmidt2007,ananth2010, lu2017} The only difference among these approaches arises from the treatment of the electronic potential term $\text{Tr}_\mathrm{e}\prod_{\alpha = 1}^{N}[e^{-\beta_{N}\hat {H}_{e}({\bf R_{\alpha}})}]$. For example, in the mean-field RPMD approach,\cite{hele2011,duke2016} the electronic potential is obtained from a weighted average of ring-polymer in different electronic configurations; in the KC-RPMD approach,\cite{menzeleev2014} the potential is obtained from the averaged ring-polymer kink configurations; in the MV-RPMD\cite{ananth2013} approach, the electronic states are explicitly described with mapping variables in the Wigner representation;\cite{hele2016} in the NRPMD\cite{richardson2013} approach, the electronic states are described with mapping variables in both position and momentum bases.

\subsection{Mapping representation for electronic states} 
We use Meyer-Miller-Stock-Thoss (MMST)\cite{meyer1979,stock1997,stock1999} mapping representation to transform the discrete electronic states into continuous variables. Based on this representation, $L$ diabatic electronic states are mapped onto $L$ harmonic oscillators' ground and first excited states with the following relation $| n \rangle \rightarrow |0_{1}...1_{n}...0_{L} \rangle=\hat{a}^{\dagger}_{n}|0_{1}...0_{n}...0_{L} \rangle$. Here, $|n\rangle$ is the diabatic state, and $|0_{1}...1_{n}...0_{L} \rangle$ is the singly excited oscillator (SEO) state with $L-1$ oscillators in their ground states and the $n_\mathrm{th}$ oscillator in its first excited state. Thus, MMST formulation provides the following mapping relation $| n \rangle \langle m| \rightarrow\hat{a}^{\dagger}_{n}\hat{a}_{m}$, with $\hat{a}^{\dagger}_{n}={1/{\sqrt{2\hbar}}}\left(\hat{q}_{n}-\mathrm{i}\hat{p}_{n}\right)$ and $\hat{a}_{m}={1/{\sqrt{2\hbar}}}\left(\hat{q}_{m}+\mathrm{i}\hat{p}_{m}\right)$ as the creation and annihilation operators for harmonic oscillator. With MMST mapping representation, the state-dependent potential operator in Eqn.~\ref{eqn:tot-H} is transformed to
\begin{equation}
\sum_{n,m}V_{nm}({\bf R_\alpha})|n\rangle\langle m| \rightarrow \sum_{n,m} V_{nm}({\bf R_\alpha})\hat a_{n}^{\dagger}\hat a_{m}.
\end{equation}

Using the above mapping relation, we can rewrite the partition function as
 \begin{eqnarray}\label{eqn:mapz}
 \mathcal{Z} &=& \lim_{N\to\infty}   \int d \{{\bf P_\alpha}\} d\{{\bf R_\alpha}\} e^{-\beta_{N}H_\mathrm{rp}} \\
&& \times\text{Tr}_\mathrm{e} \prod_{\alpha = 1}^{N} \left[e^{-\beta_{N} \sum_{nm}V_{nm}({\bf R_\alpha})\hat a_{n}^{\dagger}\hat a_{m}}\right].\nonumber
 \end{eqnarray}
To proceed, we need to choose a convenient basis to evaluate the operators $\hat a_{n}^{\dagger}$ and $\hat a_{m}$ inside $\mathcal{Z}$. Recall that coherent states  $|{\bf p,q}\rangle= |p_{1}q_{1},...,p_{n}q_{n},...p_{L}q_{L}\rangle$ are the eigenstates of the creation and annihilation operators, with the following eigen equations
 \begin{equation}\label{eqn:coh}
 \hat{a}_m |{\bf p,q}\rangle = {{(q_m + ip_m)}\over {\sqrt{2\hbar}}} |{\bf p,q}\rangle;~\langle {\bf p,q}  | \hat{a}_n^{\dagger} = \langle {\bf p,q} | {{(q_n - ip_n)}\over {\sqrt{2\hbar}}},
 \end{equation} 
where ${\bf q} \equiv \{q_{1},...q_{n},...q_{L}\}$ and ${\bf p} \equiv \{p_{1},...p_{n},...p_{L}\}$. 
 
The overlap between the coherent state basis and the diabatic basis can be expressed as follows
\begin{eqnarray}
\langle {\bf p,q}|n\rangle &=&\langle {\bf p,q}|0_{1}...1_{n}...0_{L}\rangle= {{(q_n-ip_n)} \over {\sqrt {2 \hbar}}}e^{-({\bf q}^\mathrm{T}{\bf q}+{\bf p}^\mathrm{T}{\bf p})/4\hbar} \nonumber \\
\\
\langle m | {\bf p,q}\rangle&=&\langle 0_{1}...1_{m}...0_{L}| {\bf p,q}\rangle= {{(q_m+ip_m)} \over {\sqrt {2 \hbar}}}e^{-({\bf q}^\mathrm{T}{\bf q}+{\bf p}^\mathrm{T}{\bf p})/4\hbar} \nonumber
\end{eqnarray}

\subsection{Derivation of the CS-RPMD Hamiltonian}
Expanding the exponential of electronic Hamiltonian operator in Eqn.~\ref{eqn:mapz} up to the linear order of $\beta_{N}$, under the limit that $\beta_{N}\rightarrow 0$, we obtain an equivalent expression as follows
  \begin{eqnarray}\label{eqn:hightz}
  \mathcal{Z} &=&\lim_{N\to\infty}   \int d \{{\bf P_\alpha}\} d\{{\bf R_\alpha}\} e^{-\beta_{N}H_\mathrm{rp}} \\
        &&\times\mathrm{Tr_{e}} \prod_{\alpha = 1}^{N}  \left[1-\beta_{N} \sum_{nm}V_{nm}({\bf R_\alpha})\hat a_{n}^{\dagger}\hat a_{m}+ \mathcal{O}(\beta^{2}_{N}) \right] \nonumber
 \end{eqnarray}

To proceed, recall the commutation relationship between the creation and annihilation operators $\hat a_{n}^{\dagger} \hat a_{m} =  \hat{a}_{m} \hat a_{n}^{\dagger} - \delta_{nm}$. Using this relation,  Eqn.~\ref{eqn:hightz} becomes
 \begin{eqnarray}\label{eqn:hightz2}
  &&\mathcal{Z} = \lim_{N\to\infty}   \int d \{{\bf P_\alpha}\} d\{{\bf R_\alpha}\} e^{-\beta_{N}H_\mathrm{rp}} \\
  &&\times \mathrm{Tr_{e}} \prod_{\alpha = 1}^{N} \left[1-\beta_{N} \sum_{nm}V_{nm}({\bf R_\alpha})(\hat a_{m}\hat a_{n}^{\dagger}-\delta_{nm})+\mathcal{O}(\beta^{2}_{N})\right]. \nonumber
 \end{eqnarray}

Now by inserting $N$ copies of the resolution of identity for coherent state $I_{\bf p,q} = (1/2\pi \hbar)^{L}\int d{\bf p}_{\alpha} d{\bf q}_{\alpha} | {\bf p_\alpha},{\bf q_\alpha} \rangle \langle {\bf p_\alpha} , {\bf q_\alpha} |$ in Eqn.~\ref{eqn:hightz2}, and leaving out the higher order terms $\mathcal{O}(\beta^{2}_{N})$ under $\beta_{N}\rightarrow 0$ limit, we have
\begin{eqnarray}\label{eqn:hightz3}
  &&\mathcal{Z} \propto \lim_{N\to\infty} \int d \{{\bf P_\alpha}\} d\{{\bf R_\alpha}\} e^{-\beta_{N}H_\mathrm{rp}} \int  d\{{\bf p_\alpha}\} d\{{\bf q_\alpha\}} \\ 
  &&\times \mathrm{Tr_{e}}\prod_{\alpha = 1}^{N}  \bigg[ |{\bf p_\alpha q_\alpha} \rangle \langle {\bf p_\alpha q_\alpha}|-\beta_{N} \sum_{nm}V_{nm}({\bf R_\alpha})\bigg(\hat{a}_{m} |{\bf p_\alpha q_\alpha} \rangle \langle {\bf p_\alpha q_\alpha}| \hat{a}_{n}^{\dagger}\nonumber\\
 &&~~~~~~~~-\delta_{nm} |{\bf p_\alpha q_\alpha} \rangle \langle {\bf p_\alpha q_\alpha}|\bigg)\bigg],\nonumber
 \end{eqnarray}
with $d\{{\bf p_\alpha}\} d\{{\bf q_\alpha\}}=\prod_{\alpha = 1}^{N}d{\bf p_\alpha} d{\bf q_\alpha}$

Further applying Eqn.~\ref{eqn:coh} to evaluate $\hat{a}_{m}$ and $\hat{a}_{n}^{\dagger}$, setting $\hbar=1$ from now on, and inserting the diabatic projection operator $\mathcal{P}=\sum_{n}|n\rangle\langle n|$ in-between the coherent state basis to ensure correct projection onto the finite subspace of SEOs,\cite{ananth2010,ananth2013} we obtain the following expression 
\begin{eqnarray}\label{eqn:cohz}
&&\mathcal{Z} \propto \lim_{N\to\infty} \int d \{{\bf P_\alpha}\} d\{{\bf R_\alpha}\} e^{-\beta_{N}H_\mathrm{rp}} \int  d\{{\bf p_\alpha}\} d\{{\bf q_\alpha\}}  \\
&&\times\prod_{\alpha = 1}^{N} \langle {\bf p_\alpha} , {\bf q_\alpha} | \sum_{n}|n\rangle \langle n | {\bf p_{\alpha +1}},{\bf q_{\alpha+1}}\rangle  \nonumber\\ 
&&\times \left\{1-\beta_{N} \sum_{nm}V_{nm}({\bf R_\alpha}) \left[\frac {1}{2}[{\bf q_{\alpha}}+i{\bf p_{\alpha}}]_{m} [{\bf q_{\alpha}}-i{\bf p_{\alpha}}]_{n}-\delta_{nm}\right]\right\}. \nonumber
 \end{eqnarray} 

Based on a similar derivation procedure developed for the real-time propagator,\cite{hsieh2012,hsieh2013} now we express the third line of the above expression back to the full exponential factor, and explicitly evaluate the overlap between the coherent state basis and the diabatic basis. We arrive at the final expression of the coherent state partition function $\mathcal{Z}_\mathrm{cs}$ as the central result of this paper
\begin{eqnarray}\label{eqn:csmz}
\mathcal{Z}_\mathrm{cs} &\propto& \lim_{N\to\infty} \int d \{{\bf P_\alpha}\} d\{{\bf R_\alpha}\} e^{-\beta_{N}H_\mathrm{rp}}  \int  d\{{\bf p_\alpha}\} d\{{\bf q_\alpha\}} \\
&&\times \prod_{\alpha = 1}^{N} \frac{1}{2}({\bf q_{\alpha}} -i{\bf p_{\alpha}})^\mathrm{T} ({\bf q_{\alpha+1}} +i{\bf p_{\alpha+1}}) e^{-{1\over 2}({\bf q}^\mathrm{T}_{\alpha}{\bf q}_{\alpha}+{\bf p}^\mathrm{T}_{\alpha}{\bf p}_{\alpha})} \nonumber \\ 
&&\times e^{-\beta_{N} \sum_{nm}V_{nm}({\bf R_{\alpha}})\left[\frac{1}{2}([{\bf q_{\alpha}}]_{m} [{\bf q_{\alpha}}]_{n} + [{\bf p_{\alpha}}]_{m} [{\bf p_{\alpha}}]_{n})-\delta_{nm}\right]}.  \nonumber
 \end{eqnarray} 
 
The above coherent state partition function can be written into more compact form as follows
\begin{equation}\label{eqn:csmz-ham}
\mathcal{Z}_\mathrm{cs} \propto \lim_{N\to\infty} \int d \{{\bf P_\alpha}\} d\{{\bf R_\alpha}\} \int d\{{\bf p_\alpha}\} d\{{\bf q_\alpha\}} {\bf \Gamma} e^{-\beta_{N}H_\mathrm{cs}},    
\end{equation} 
with the weighting factor 
\begin{equation}\label{eqn:gamma}
{\bf \Gamma}= \prod_{\alpha = 1}^{N}\frac{1}{2}({\bf q_{\alpha}} -i{\bf p_{\alpha}})^\mathrm{T} ({\bf q}_{\alpha+1} +i{\bf p}_{\alpha+1})e^{-{1\over 2}({\bf q}^\mathrm{T}_{\alpha}{\bf q}_{\alpha}+{\bf p}^\mathrm{T}_{\alpha}{\bf p}_{\alpha})},
\end{equation}
and the CS-RPMD Hamiltonian  
\begin{eqnarray}\label{eqn:csham}
H_\mathrm{cs}&=& \sum_{\alpha=1}^{N}{{{\bf P_{\alpha}}^{2}}\over{2M}}+V_{0}({\bf R_{\alpha}})+{M\over{2\beta^{2}_{N}\hbar^{2}}}({\bf R_{\alpha}-R_{\alpha-1}})^{2} \\
&+& \sum_{nm}V_{nm}({\bf R_{\alpha}})\left[\frac{1}{2}([{\bf q_{\alpha}}]_{m} [{\bf q_{\alpha}}]_{n} + [{\bf p_{\alpha}}]_{m} [{\bf {p_{\alpha}}}]_{n})-\delta_{nm}\right].\nonumber
\end{eqnarray}

The first line in $H_\mathrm{cs}$ corresponds to the nuclear ring-polymer part of the Hamiltonian, and the second line corresponds to the non-adiabatic part of the Hamiltonian which describes electrons-nuclei interactions. Note that $H_\mathrm{cs}$ does not contain a potential that couples two adjacent mapping beads. Similar feature has also been proposed in the NRPMD Hamiltonian,\cite{richardson2013} which ensures to capture the correct electronic Rabi oscillations. Here, we derived $H_\mathrm{cs}$ with this feature. To be specific, when the electronic DOF is decoupled from nuclear DOF, such that $V_{nm}({\bf R_{\alpha}})=V_{nm}$, the non-adiabatic part of $H_\mathrm{cs}$ becomes MMST Hamiltonian with bead-averaged initial conditions, which gives exact frequency for electronic Rabi oscillations.\cite{stock1997,stock1999} Meanwhile, MV-RPMD\cite{ananth2013} does contain inter-bead coupling for mapping DOF and cannot capture the correct Rabi oscillations.\cite{althorpe2016} 

We would like to emphasize several other key features of CS-RPMD approach. First, CS-RPMD will reduce back to regular adiabatic RPMD with decoupled electrons-nuclei limit, including state-independent limit $V_{nm}=0$ ($n\neq m$) and the adiabatic limit $\beta V_{nm}\gg 1$. Second, CS-RPMD has a very clear one-bead limit. With only one bead for both mapping and nuclear DOFs, CS-RPMD Hamiltonian reduces back to the MMST mapping Hamiltonian in the coherent state representation. Third, through a wick rotation, $\beta\rightarrow it/\hbar$, coherent state partition function $Z_\mathrm{cs}$ in Eqn.~\ref{eqn:csmz} becomes a coherent state real-time propagator used in the Forward-Backward MQCL (FB-MQCL)\cite{hsieh2012,hsieh2013} and the Partial-Linearized Density Matrix (PLDM) approach.\cite{huo2011,huo2013} In FB-MQCL, the ${\bf \Gamma}$ term appears as the overlap between two coherent state bases from two consecutive real-time propagators.\cite{hsieh2012,hsieh2013}

We further emphasize that the possible mapping RPMD Hamiltonian expression is not unique, for instance, we have also obtained the MV-RPMD Hamiltonian in the coherent state representation as presented in Appendix A. The reason for this non uniqueness is that the quantum partition function in Eqn.~\ref{eqn:part-gen} is an integral of the function of Hamiltonian, thus it is mathematically possible to find different Hamiltonian (integrand) that gives the same quantum partition function (integral). 

\subsection{CS-RPMD Time Correlation Function}
With the derived partition function (Eqn.~\ref{eqn:csmz-ham}) and $H_\mathrm{cs}$ (Eqn.~\ref{eqn:csham}), we {\it propose} to use them to compute the Kubo-transformed time correlation function for operators $\hat{A}$ and $\hat{B}$
\begin{equation}\label{eqn:kubo}
\tilde{C}_{AB}(t) = \frac{1}{\mathcal{Z}\beta} \int_{0}^{\beta} \mathrm{Tr}[e^{-(\beta - \lambda)\hat H}\hat A e^{-\lambda \hat H}e^{i\hat H t/ \hbar} \hat B e^{-i\hat H t/ \hbar}]d\lambda.
\end{equation}
Similar to the original RPMD approach, we propose that the CS-RPMD correlation function
\begin{eqnarray}\label{eqn:cs_corr}
 C_{AB}(t) &=& \frac{1}{\mathcal{Z}_\mathrm{cs}} \int d\{{\bf R_\alpha}\} d\{{\bf P_\alpha}\}d\{{\bf q_\alpha}\}d\{{\bf p_\alpha}\} \\
              &&~~~~~~\times {\bf\Gamma}(0)e^{-\beta_{N}\hat H_{\mathrm{cs}}}\bar{A}(0)\bar{B}(t),\nonumber
\end{eqnarray}
is an approximate Kubo-transformed time correlation function,\cite{habershon2013, hele2016} where $\bar{A}(0)$ and $\bar{B}(t)$ are the bead-averaged estimators for the corresponding operators. Note that ${\bf\Gamma}(0)$ is Eqn.~\ref{eqn:gamma} evaluated with the initial mapping variables at t=0 , and $\bar{B}(t)$ is evaluated with the classical trajectories generated from the Hamiltonian $H_\mathrm{cs}$. In this paper, we are interested in two types of auto-correlation functions: (1) nuclear position auto-correlation function $C_{RR}(t)$ where $\hat{A}=\hat{B}=\hat{R}$, thus $\bar{A}=\bar{B}=\frac{1}{N}\sum_{\alpha=1}^{N}{\bf R}_\alpha$, and (2) population auto-correlation function $C_{nn}(t)$ where $\hat{A}=\hat{B}=|n\rangle\langle n|$ with the corresponding estimator 
\begin{equation}\label{eqn:p-est}
\bar{A}=\bar{B}=\frac{1}{N}\sum_{\alpha=1}^{N}{{[{\bf q}_{\alpha} -i{\bf p}_{\alpha}]_{n} [{\bf q}_{\alpha+1} +i{\bf p}_{\alpha+1}]_{n}}\over{({\bf q}_{\alpha} -i{\bf p}_{\alpha})^\mathrm{T} ({\bf q}_{\alpha+1} +i{\bf p}_{\alpha+1})}}.
\end{equation}

The CS-RPMD time correlation function can be computed by sampling initial configurations with NVT trajectories generated from $H_\mathrm{cs}$, then propagating the dynamics with same Hamiltonian $H_{cs}$ to evaluate $\bar{B}(t)$. Each trajectory is weighted by an {\it initial} weighting factor ${\bf \Gamma}$(Eqn.~\ref{eqn:gamma}).  Note that both ${\bf \Gamma}$ as well as the projection operator estimator are complex. We thus use their complex values to accumulate the time correlation function. We obtain results of ${C}_{AB}(t)$ with zero complex values within numerical errors.

The novelty of the CS-RPMD formalism is that through the classical evolution of $H_\mathrm{cs}$, the initial configuration governed by $e^{-\beta_{N}H_\mathrm{cs}}$ is preserved for the ensemble of trajectories. Thus, CS-RPMD provides a stable propagation scheme for the dynamics and avoids initial configuration leakage problems. NRPMD approach,\cite{richardson2013} on the other hand, uses one Hamiltonian (that closely resembles MV-RPMD Hamiltonian) to sample the initial configurations and then another Hamiltonian (that closely resembles CS-RPMD Hamiltonian) to propagate dynamics, and thus might encounter initial configuration leakage problem. Similarly,  in commonly used linearized path-integral approaches (classical Wigner methods),\cite{sun1998,sun1998,shi2004,huo2011,huo2013} the classical nuclear propagation cannot preserve Wigner initial distribution, and might cause zero-point energy leakage problem.\cite{habershon2009} Just like any imaginary-time path-integral method, CS-RPMD uses classical dynamics in the extended phase space to quantize nuclei rather than initially enforces ZPE through Wigner distribution, thus significantly alleviating the ZPE leakage problem encountered in the classical Wigner methods.\cite{habershon2009} 

However, CS-RPMD does not preserve QBD in general, despite the fact that the partition function in Eqn.~\ref{eqn:csmz-ham} exactly describes the QBD for any quantum statistics calculations (under the $N\to\infty$ limit for both nuclear and mapping variables). As can be clearly seen in Eqn.~\ref{eqn:csmz-ham}, CS-RPMD partition function requires an additional weighting factor ${\bf \Gamma}$ to be multiplied with $e^{-\beta_{N}H_{cs}}$ in order to recover QBD. This can be easily accomplished for quantum statistics calculations. On the other hand, for any time correlation function calculation, the pre-factor ${\bf \Gamma}$ (Eqn.~\ref{eqn:gamma}) is not a constant during the classical evolution governed by $H_\mathrm{cs}$. Thus, CS-RPMD time correlation function calculations which only use ${\bf \Gamma}(0)$ as an {\it initial} weighting factor, does not preserve QBD in general, especially at the longer time. As a consequence, CS-RPMD in general does not preserve ZPE associated with QBD, except at time $t=0$, or under the adiabatic or state-independent limit. Despite this deficiency, our numerical results demonstrate that accurate time correlation function can still be obtained. In the state-independent or the adiabatic limit, on the other hand, CS-RPMD reduces back to regular RPMD and preserves QBD, just like any state-dependent RPMD approach.\cite{ananth2013} 

For a special case where there is only one bead for the mapping DOF (such that all mapping beads collapse into one), but still multiple beads for the nuclear DOF, ${\bf \Gamma}={1\over2}\sum_{n}({q}_{n}^{2}+{p}_{n}^{2})e^{-{1\over 2}({\bf q}^\mathrm{T}{\bf q}+{\bf p}^\mathrm{T}{\bf p})}=\langle {\bf pq}|\sum_{n}|n\rangle\langle n|{\bf pq} \rangle$ is indeed the integral of motion of $H_\mathrm{cs}$. Thus in this special case, CS-RPMD preserves detailed balance, such that\cite{hele2015,liu2011} $C_{AB}(t)=C_{BA} (-t)$, with an approximate QBD generated from one mapping bead and $N$ nuclear beads. Under this limit,  CS-RPMD Hamiltonian in Eqn.~\ref{eqn:csham} becomes $H_\mathrm{cs}=\sum_{\alpha=1}^{N}{{{\bf P_{\alpha}}^{2}}\over{2M}}+V_{0}({\bf R_{\alpha}})+{M\over{2\beta^{2}_{N}\hbar^{2}}}({\bf R_{\alpha}-R_{\alpha-1}})^{2}+ \sum_{nm}V_{nm}({\bf R_{\alpha}})\left[\frac{1}{2}(q_{m}q_{n} + p_{m}p_{n})-\delta_{nm}\right]$, which is the ring-polymer nuclei under the coherent state MMST potential. Note that the projection operator $\mathcal{P}=\sum_{n}|n\rangle\langle n|$ in the $\bf \Gamma$ expression constrains the electronic mapping variables within the physical SEO subspace.\cite{ananth2010} This gives the partition function $\mathrm{Tr}[{\bf \Gamma}e^{-\beta H_\mathrm {cs}}]$ (with $\mathrm{Tr}\equiv\int {\bf dR dq dp}$) rather than simply assuming a classical partition function $\mathrm{Tr}[e^{-\beta H_\mathrm {cs}}]$.

In addition, CS-RPMD preserves Rabi oscillations as well as detailed balance under this special one-bead mapping limit. However, we need to emphasize that by using only one bead for the mapping variables, one cannot fully recover the exact QBD even at t=0. In order to obtain the exact QBD (see Eqn.~\ref{eqn:hightz3}), we explicitly require $\beta_N\rightarrow 0$ (and $N\rightarrow \infty$) for both electronic and nuclear DOFs in our derivations. Nevertheless, our numerical results (provided in Appendix C) suggest that the quantum statistics obtained from this approximated partition function is close to the exact QBD, and the time correlation function is close to both exact results and the CS-RPMD calculation with multiple beads for all DOFs. We want to further emphasize that this limiting case has already proven to be useful in recently developed approaches, such as Ehrenfest RPMD\cite{yoshikawa2013} and RPSH\cite{shushkov2012,shakib2017} which assume one bead for the electronic DOF and multiple beads for the nuclear DOF.

Finally, due to the presence of MMST mapping Hamiltonian in $H_\mathrm{cs}$, extra caution is still needed. It is well known that this Hamiltonian could exhibit the ``inverted potential" problem\cite{coker2001} when ${1\over2}([q_{\alpha}]^{2}_{n}+[p_{\alpha}]^{2}_{n})-1<0$, as well as the ZPE leakage problem associated with the mapping DOF due to the explicitly incorporated ZPE term, {\it i.e.}, $-\sum_{n}V_{nn}({\bf R}_\alpha)$. These intrinsic deficiencies associated with the MMST Hamiltonian can be addressed and refined with the recent theoretical developments, such as applying semi-classical approximation to treat the mapping DOF,\cite{bonella2003,huo2011} using ZPE corrections for the mapping variables,\cite{stock1999,miller2016} or using new forms of mapping Hamiltonians.\cite{miller2015,liu2016}

\subsection{Connections and Differences with Other Methods}
CS-RPMD Hamiltonian is derived with the coherent state basis for mapping variables, and it is closely related to two recently developed state-dependent RPMD formalisms, MV-RPMD\cite{ananth2013} and NRPMD.\cite{richardson2013}  Here we want to clarify the connections and differences between CS-RPMD and these two methods. 

First, MV-RPMD Hamiltonian does contain inter-bead couplings for the mapping DOF,\cite{ananth2013} and cannot correctly capture the electronic Rabi oscillations in a bare two state system.\cite{althorpe2016} In addition, the inter-bead couplings for mapping DOF in MV-RPMD might cause unphysical oscillation frequency even in the nuclear auto-correlation function calculations, as demonstrated in the result section. We further emphasize that the possible form of the state-dependent Hamiltonian that gives exact QBD for quantum statistics calculations is not unique. For instance, we can also obtain the MV-RPMD Hamiltonian\cite{ananth2013} in the coherent state representation, as presented in Appendix A. On the other hand, MV-RPMD approach does preserve QBD throughout its dynamical propagation, which is an appealing feature.

Second, the NRPMD approach\cite{richardson2013} uses a proposed Hamiltonian that closely resembles $H_\mathrm{cs}$ (Eqn.~\ref{eqn:csham}) to propagate dynamics, thus provides accurate electronic Rabi oscillations. However, the initial phase space points in this method are not sampled from the same Hamiltonian, but from another one that contains inter-bead coupling terms in the mapping potential.\cite{richardson2013} By doing so, each trajectory in the NRPMD approach might move out of the original phase space distribution due to using two different Hamiltonians. Thus, NRPMD might encounter initial configuration leakage problem. In the coherent state basis context, this corresponds to a method that uses the coherent state MV-RPMD Hamiltonian $H_\mathrm{cmv}$ (Eqn.~\ref{eqn:mv-rpmd} in Appendix A) to sample initial configurations, and then uses CS-RPMD Hamiltonian, $H_\mathrm{cs}$ (Eqn.~\ref{eqn:csham}) to propagate trajectories. Besides that, NRPMD also requires two estimators for projection operator in order to efficiently compute the population correlation function.\cite{richardson2013,richardson2017} CS-RPMD approach, on the other hand, uses only one estimator (Eqn.~\ref{eqn:p-est}).

Compared to these two recently developed mapping RPMD approaches, the merits of CS-RPMD are (1) preserving the electronic Rabi oscillations that MV-RPMD fails to describe, through a rigorously derived $H_\mathrm{cs}$ that does not contain any inter-bead coupling terms for mapping DOF, (2) sampling and propagating trajectories from one {\it derived} Hamiltonian, as opposed to NRPMD approach that uses two Hamiltonians without rigorous derivations, and (3) preserving detailed balance with an approximate QBD in a special case which contains only one mapping bead. 

However, we must admit that the numerical results obtained with CS-RPMD (for the current model systems) are not significantly different than those obtained from NRPMD. This suggests that both $H_\mathrm{cs}$ and the Hamiltonian proposed in NRPMD (which closely resembles $H_\mathrm{cs}$) provide accurate dynamics. While the the NRPMD approach simply assumes a Hamiltonian of this form,\cite{richardson2013} the current work demonstrates how to rigorously derive this Hamiltonian from a partition function that provides exact QBD.

\section{Results and Discussion.} To test the accuracy of CS-RPMD, we adapt a commonly used model system that contains one nuclear coordinate and two electronic states\cite{ananth2013,richardson2013} 
 \begin{equation}
 \hat H = \frac{\hat P^2}{2M} + \frac{1}{2}M\omega^2 \hat R^2 + \begin{bmatrix}\epsilon + k\hat R & \Delta \\ \Delta & - \epsilon - k \hat R \end{bmatrix},
 \end{equation}
where $\Delta$ is the electronic coupling, $k$ is the vibronic coupling, and 2$\epsilon$ is the energy bias between the two diabatic states. In this paper, we choose a reduced unit system such that  $M=\hbar = 1$ and $\omega= \beta = k= 1$. 

Table I presents the parameters for all of the model systems used in this paper. In particular, Model I and V are in the adiabatic regime, where $\Delta\gg \beta^{-1}$; Model II and III are in the non-adiabatic regime, where $\Delta\ll \beta^{-1}$; Model IV and VI are in the intermediate regime, where $\Delta\sim \beta^{-1}$. Model III and VI are asymmetric cases with finite diabatic energy bias $2\epsilon$, and the rest of the model systems are symmetric cases with $\epsilon$=0.
\begin{table}[h]
\begin{tabular}{ p{1.0cm} p{0.8cm} p{0.8cm} p{0.8cm}  p{0.8cm} p{0.8cm} p{0.8cm}}
\hline
\hline
 & I & II & III & IV & V & VI  \\
\hline
$\epsilon$ & 0  & 0 & 1.5 & 0 & 0 & 2\\
$\Delta$   &10   & 0.10 & 0.10 & 1 & 4 & 1\\
  \hline
 \hline
\end{tabular}
\caption{Parameters (in a.u.) for model systems I-VI .}
\end{table}
\begin{figure}
 \centering
  \begin{minipage}[t]{1.0\linewidth}
     \centering
     \includegraphics[width=\linewidth]{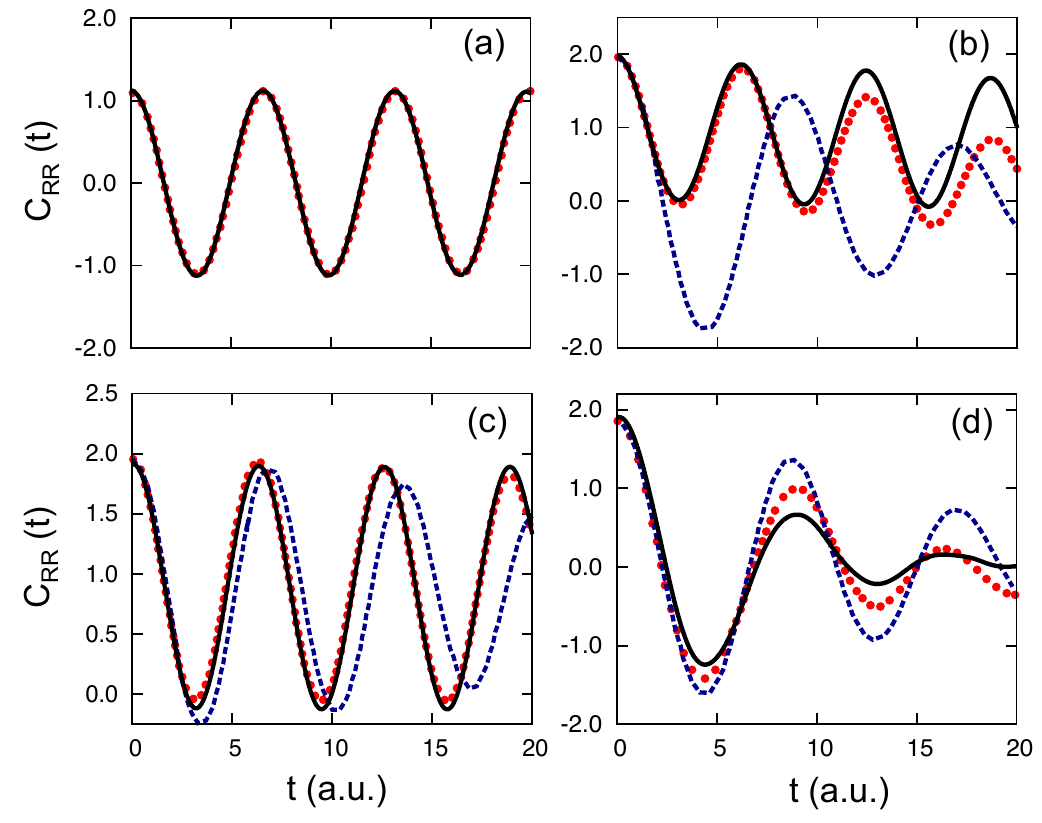}
  \end{minipage}%
   \caption{The Kubo-transformed nuclear position auto-correlation function for model I-IV obtained from CS-RPMD (black solid), the mean-field RPMD (blue dash), and numerical exact results (red dot). Results for Model I (symmetric, adiabatic) are in panel (a), Model II (symmetric, non-adiabatic) are in panel (b), Model III (asymmetric, non-adiabatic) are in panel (c), and Model IV (symmetric, intermediate) are in panel (d). }
\label{fig:rrcorr}
\end{figure}

CS-RPMD correlation functions are computed from Eqn.~\ref{eqn:cs_corr}. To evaluate the ensemble average, a total number of $10^{4}$ initial configurations are sampled from a $2\times10^{7}$ a.u. long CS-RPMD NVT trajectory, thermostatted by resampling the nuclear velocities from the Maxwell-Boltzmann distribution at every $0.4/\Delta$ a.u. Each configuration is then further equilibrated with NVE CS-RPMD propagation for another 200 a.u., before being used to propagate and accumulate the CS-RPMD correlation function. All of the correlation functions converge at $N=8$ or fewer beads. Numerical exact results are obtained from discrete variable representation (DVR) calculations. \cite{colbert1992}

Figure~\ref{fig:rrcorr} presents nuclear position auto-correlation function computed from CS-RPMD (black),  the mean-field RPMD\cite{hele2011,ananth2013} approach (blue) with expression provided in Appendix B, and numerical exact method (red) for Models I-IV. Model I in Fig.~\ref{fig:rrcorr}a is in the adiabatic regime. In this case, CS-RPMD goes back to the standard RPMD, and agrees with the exact result due to the near Harmonic adiabatic potential. Mean-field RPMD in this case also gives the same exact result thus not shown here. Model II in Fig.~\ref{fig:rrcorr}b  is in the non-adiabatic regime. This is the most challenging case and the most relevant regime for non-adiabatic electron transfer\cite{menzeelev2011} and proton-coupled electron transfer reactions.\cite{kretchmer2013} In this regime, mean field RPMD starts to break down even at a very short time. CS-RPMD, on the other hand, performs reasonably well compared to exact DVR calculations at the longer time. Models III and IV are in the intermediate regime, with asymmetric (Fig.~\ref{fig:rrcorr}c) and symmetric (Fig.~\ref{fig:rrcorr}d) diabatic bias $2\epsilon$. In this regime, both CS-RPMD and mean field RPMD behave reasonably well. 

We have also tested the special case when there is only one bead for mapping DOF, with the results provided in Appendix C. With $N=8$ beads for the nuclear DOF and only one bead for the mapping DOF, we found good agreements for these correlation functions compared to the results obtained with multiple beads for all DOFs, suggesting a promising practical strategy for preserving detailed balance as well as providing accurate non-adiabatic dynamics. 

\begin{figure}
 \centering
  \begin{minipage}[t]{0.95\linewidth}
     \centering
     \includegraphics[width=\linewidth]{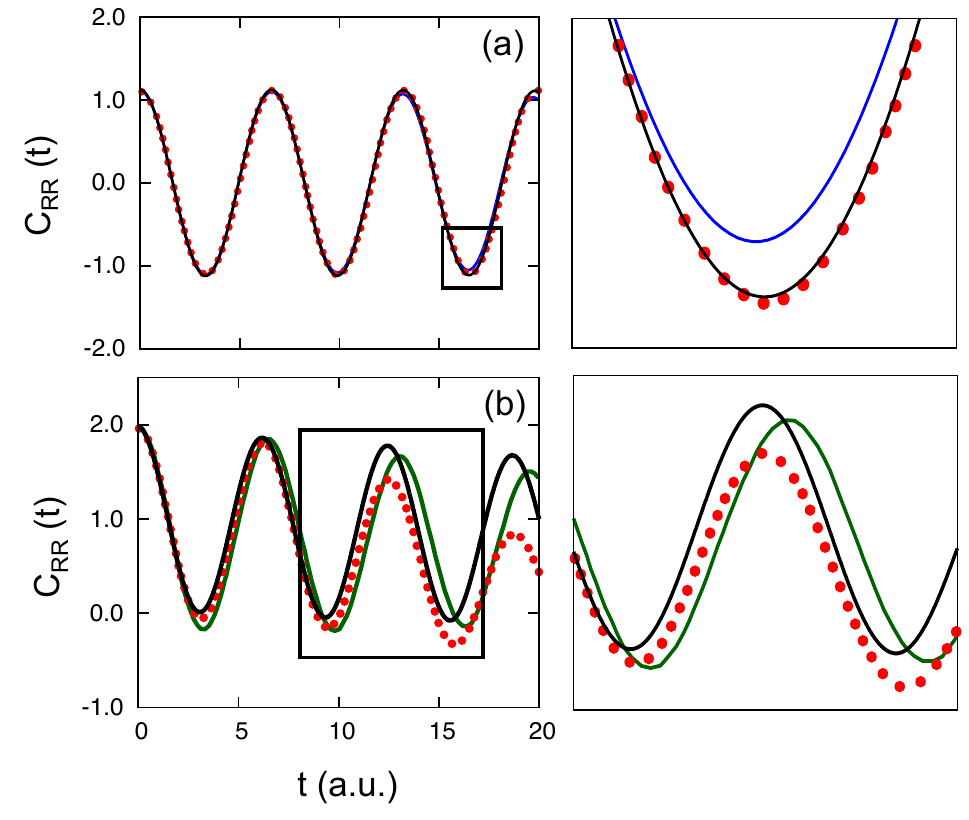}
  \end{minipage}%
   \caption{Comparison between CS-RPMD with NRPMD\cite{richardson2013} and MV-RPMD.\cite{ananth2013} (a)  The Kubo-transformed nuclear position auto-correlation function for Model I obtained from CS-RPMD (black), NRPMD (blue), and numerical exact results (red). In this adiabatic case, MV-RPMD agrees with numerical exact result, and thus not shown. (b) Results for Model II obtained from CS-RPMD (black), MV-RPMD (green), and numerical exact results (red). In this non-adiabatic case, NRPMD result is close to CS-RPMD, and thus not shown. Magnified plots that correspond to the square regions in both panels are provided on the right hand side. }
\label{fig:comp}
\end{figure}

Figure~\ref{fig:comp} presents the comparison between CS-RPMD, NRPMD, and MV-RPMD. In Figure~\ref{fig:comp}(a), Model I is used to illustrate the initial distribution leakage problem that NRPMD encounters. Here we present the results obtained from NRPMD\cite{richardson2013} (blue), CS-RPMD (black) and numerical exact method (red). In this adiabatic test case, all of the state-dependent RPMD approaches are expected to reduce to the regular RPMD approach, and reproduce the exact result due to the nearly harmonic adiabatic potential in this model. This is indeed the case for MF-RPMD\cite{ananth2013}, MV-RPMD\cite{ananth2013}, and CS-RPMD, all of which can fully recover the exact result under this limit. However, as can be seen from Figure~\ref{fig:comp}(a), the result from NRPMD approach starts to deviate from the exact one at longer times. This is due to the fact that in the NRPMD approach,\cite{richardson2013} the Hamiltonian used to propagate trajectories will not preserve the initial distribution for the ensemble of trajectories that is sampled from another Hamiltonian. A similar situation is also encountered in the coherent state version of NRPMD (result not shown) that samples initial configurations with $H_\mathrm{cmv}$ (Eqn.~\ref{eqn:mv-rpmd} in Appendix A) and propagates trajectories with $H_\mathrm{cs}$ (Eqn.~\ref{eqn:csham}). CS-RPMD provides accurate long-time dynamics for this adiabatic test case, by sampling and propagating trajectories with the same Hamiltonian $H_\mathrm{cs}$ and thus preserving the phase space distribution for the ensemble of trajectories governed by $H_\mathrm{cs}$ throughout the dynamical propagation. 

In Figure~\ref{fig:comp}(b), Model II is used to provide the comparison between MV-RPMD and CS-RPMD. Here we present the results obtained from MV-RPMD (green), CS-RPMD (black) and numerical exact method (red).  As can be seen, the correlation function obtained from MV-RPMD\cite{ananth2013} starts to oscillate with a different frequency compared to the quantum result at a longer time. This might happen because the inter-bead couplings for mapping DOF start to contaminate the physical frequency of the system.  A similar situation is also encountered in the coherent state MV-RPMD (result not shown) that samples and propagates trajectories with $H_\mathrm{cmv}$ (Eqn.~\ref{eqn:mv-rpmd} in Appendix A). CS-RPMD on the other hand, preserves the correct oscillation frequency in the correlation function, due to the character of $H_\mathrm{cs}$ that does not contain the inter-bead coupling for mapping DOF. NRPMD provides a similar result (not shown) compared to CS-RPMD in this non-adiabatic test case. Interestingly, NRPMD and MV-RPMD start to show problematic behaviors at the opposite limit of the electronic coupling, where CS-RPMD provides reliable results across a broad range of parameters as demonstrated in Figure~\ref{fig:rrcorr}.

\begin{figure}
 \centering
  \begin{minipage}[t]{1.0\linewidth}
     \centering
     \includegraphics[width=\linewidth]{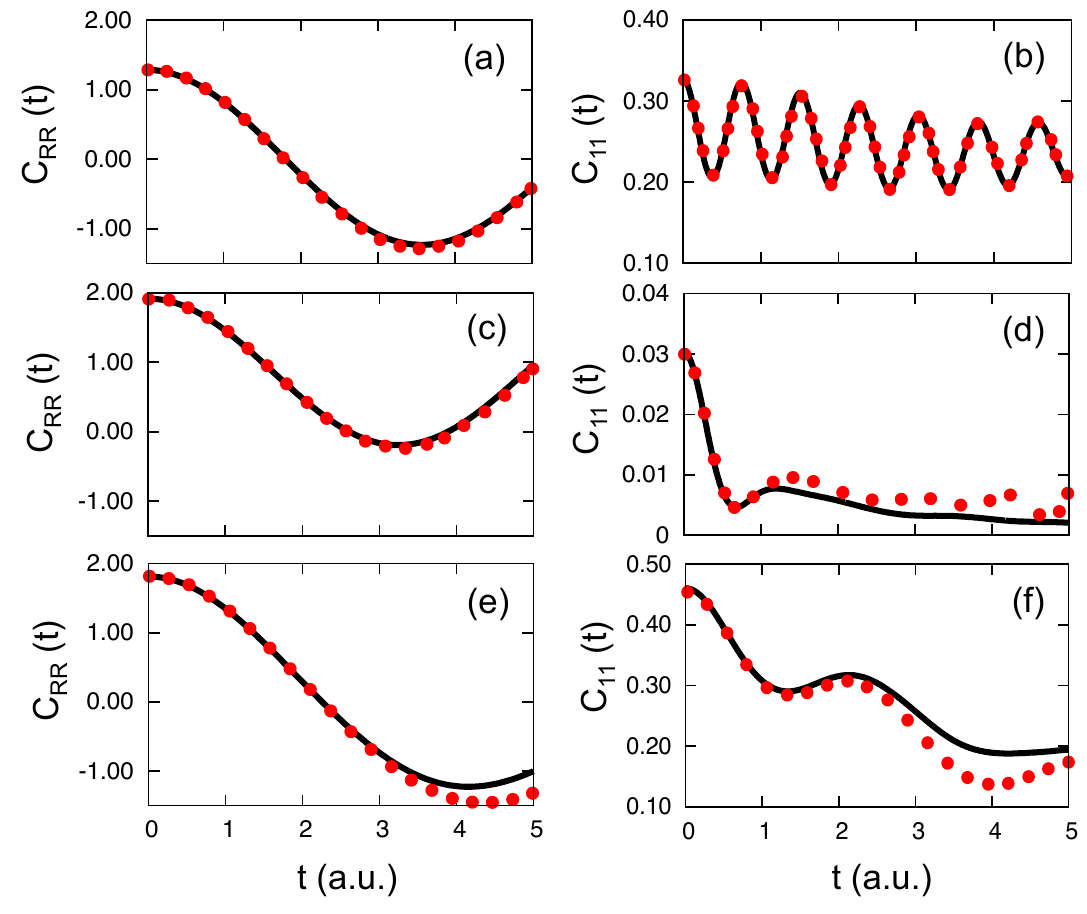}
  \end{minipage}%
   \caption{The Kubo-transformed nuclear position and the electronic population auto-correlation functions for model IV-VI obtained from CS-RPMD (black) and numerical exact DVR method (red). Results for Model V (symmetric, adiabatic) are in panel (a) and (b), model VI (asymmetric, intermediate) are in panel (c) and (d), and model IV (symmetric, intermediate) are in panel (e) and (f).}
\label{fig:popcorr}
\end{figure}

Figure~\ref{fig:popcorr} presents the nuclear position and the electronic population auto-correlation functions computed from CS-RPMD (black) and numerical exact method (red) for models IV-VI. Accurately describing electronic interference effects (Rabi oscillations) are essential for non-adiabatic dynamics simulations. Again, CS-RPMD agrees with exact results in the adiabatic regime for model V presented in Fig.~\ref{fig:popcorr}a, and provides reasonably good results for the model systems in the intermediate regimes presented in Fig.~\ref{fig:popcorr}c-f. NRPMD approach\cite{richardson2013} (results not shown) gives nearly identical results for these correlation functions. MV-RPMD on the other hand, cannot correctly capture the electronic oscillations in these population auto-correlation functions, due to the contamination of the true electronic Rabi oscillations with the inter-beads couplings in the mapping ring-polymer Hamiltonian.\cite{ananth2013,althorpe2016} 

\section{Conclusion.} In this paper, we present CS-RPMD approach, a new state-dependent RPMD method that can accurately describe electronic non-adiabatic dynamics and nuclear quantum effects. With MMST mapping representation in the coherent state basis for the electronic DOF and regular path-integral representation for the nuclear DOF, we derive the CS-RPMD Hamiltonian, which closely resembles the proposed NRPMD Hamiltonian.\cite{richardson2013,hele2016} In a special case where there is only one mapping bead (and still multiple nuclear beads), we can rigorously prove that CS-RPMD preserves detailed balance with an approximate quantum Boltzmann distribution. Numerical results from model systems demonstrate the accuracy of this approach across a broad range of electronic coupling regimes. 

Compared to recently developed state-dependent RPMD approaches\cite{richardson2013,ananth2013}, CS-RPMD provides further appealing features, including preserving the electronic Rabi oscillations that MV-RPMD\cite{ananth2013} fails to describe. In addition, compared to NRPMD approach which simply assumes a Hamiltonian of this form\cite{richardson2013}, the current work demonstrates how to derive this Hamiltonian from a partition function that contains exact QBD, providing a more solid theoretical foundation for such methods.

The equivalence between RPMD and Kubo-transformed time correlation function has been originally proposed\cite{craig2004,braams2006} and recently proved through Matsubara dynamics.\cite{hele2015,hele2015jcp2,hele2016} We envision a similar rigorous proof\cite{hele2016} of the relation between the CS-RPMD and Kubo-transformed time correlation function, with a coherent state mapping Matsubara dynamics framework in future. In addition, we envision to develop a method that exactly preserves both the quantum Boltzmann distribution and the electronic Rabi oscillations, and at the same time, scales linearly with the nuclear DOF and friendly with electronic DOF.\cite{althorpe2016} Finally, practical directions will focus on applying CS-RPMD approach to simulate the coupled electron and proton transfer reactions in large-scale condensed phase systems.\cite{menzeelev2011,kretchmer2013} 
\\
\\
{\bf Acknowledgement}\\
This work was supported by the University of Rochester startup funds. Computing resources were provided by the Center for Integrated Research Computing (CIRC) at the University of Rochester. We appreciates valuable discussions with Dr. Tim Hele and Profs. Tom Miller, Jeremy Richardson, and Nandini Ananth.  We appreciate the critical and thorough comments from both reviewers.
\\
\\
{\bf Appendix A: Derivation of the coherent-state MV-RPMD Hamiltonian.} \\
Here, we obtain MV-RPMD Hamiltonian in the coherent state mapping representation. Note that MV-RPMD is not a new method and has been derived in the Wigner basis.\cite{ananth2013} We start from the partition function expression in Eqn.~\ref{eqn:part-gen}, perform the trace over the electronic DOF with coherent state mapping basis, further insert the projection operator $\mathcal{P} = \sum_{n}|n\rangle \langle n| $, and arrive at the following expression
\begin{eqnarray}
&&\mathcal{Z} \propto \lim_{N\to\infty} \int d \{{\bf P_\alpha}\} d\{{\bf R_\alpha}\} e^{-\beta_{N}H_\mathrm{rp}}\int d\{{\bf p}_\alpha\} d\{{\bf q}_\alpha\}  \nonumber\\
&& \times\prod_{\alpha=1}^{N}  \langle {\bf p_\alpha,q_\alpha}| \sum_{n}|n\rangle \langle n|  e^{-\beta_{N}\hat{H}_\mathrm{e}({\bf R_{\alpha})}} |\sum_{m}|m\rangle \langle m|  { \bf p_{\alpha+1},q_{\alpha+1}} \rangle.\nonumber
\end{eqnarray}

The overlap between the diabatic basis and coherent state basis are explicitly evaluated, and the final expression for coherent state MV-RPMD partition function is 
\begin{equation}\label{eqn:part_mv}
\mathcal{Z}_\mathrm{cmv} \propto \lim_{N\to\infty} \int d\{{\bf R_\alpha}\} d\{{\bf P_\alpha}\}d\{{\bf q_\alpha}\} d\{{\bf p_\alpha}\} {\mathrm{sgn}(\Theta)} e^{-\beta_{N} H_\mathrm{cmv}}
\end{equation}
with the coherent state MV-RPMD Hamiltonian $H_\mathrm{cmv}$  defined as
\begin{eqnarray}\label{eqn:mv-rpmd}
H_\mathrm{cmv} &=& \sum_{\alpha=1}^{N} \left(\frac{{\bf P}_{\bf \alpha}^2}{2M} + V_{0}({\bf R_\alpha}) + \frac{M}{2\beta_{N}^2 \hbar^{2}}({\bf R_{\alpha}-R_{\alpha+1}})^{2}\right) \nonumber \\
&&+ {N\over\beta}\sum_{\alpha=1}^{N}\frac{1}{2}({\bf q}^\mathrm{T}_{\alpha}{\bf q}_{\alpha}+{\bf p}^\mathrm{T}_{\alpha}{\bf p}_{\alpha}) - \frac{N}{\beta} \ln |\Theta|.
\end{eqnarray}
Here, $\Theta = \mathrm{Re} \prod_{\alpha = 1}^{N} \sum_{nm}  [{\bf q}_{\alpha} - i {\bf p}_{\alpha}]_{n}\mathcal{M}_{nm}[{\bf q}_{\alpha+1}+i{\bf p}_{\alpha+1}]_{m}$, with $\mathcal{M}_{nm}({\bf R_{\alpha}})= \langle n |e^{-\beta_{N} V_{nm}(\bf {R}_{\alpha})} | m\rangle$. These results have been derived in the Wigner representation of the mapping variables.\cite{ananth2013}
\\
\\
{\bf Appendix B: Meanfield RPMD Hamiltonian.} \\
Here, we derive Mean-field RPMD Hamiltonian with the coherent state mapping representation. Note that MF-RPMD is not a new method and has been derived without using mapping representation.\cite{hele2011}  Our starting point is the CMV-RPMD partition function expression in Eqn.~\ref{eqn:part_mv}. The mean-field (MF) approximation is applied by integrating over the electronic mapping variables to obtain an effective potential for the nuclear DOF. Analytically evaluating the Gaussian integral over $d\{{\bf q}_{\alpha}\} d\{{\bf p}_{\alpha}\}$ in Eqn.~\ref{eqn:part_mv}, we arrived at the MF-RPMD partition function expression 
\begin{equation}
\mathcal{Z}_\mathrm{MF} \propto \lim_{N\to\infty} \int d\{{\bf P}_{\alpha}\}  d\{{\bf R}_{\alpha}\}e^{-\beta_N H_\mathrm{MF}}\mathrm{sgn}(\Theta^{\prime}),
\end{equation}
where $H_\mathrm{MF}=H_\mathrm{rp}-\frac{N}{\beta}\mathrm{ln}|\Theta^{\prime}|$ with $\Theta^{\prime} = \mathrm{Tr}_e\left[\prod_{\alpha=1}^N  \mathcal{M}({\bf R}_{\alpha}) \right]$ and $H_\mathrm{rp}$ is the regular ring-polymer Hamiltonian defined in Eqn.~\ref{eqn:hrp}.
\\
\\
{\bf Appendix C: Results for one mapping bead and multiple nuclear beads.} \\
Here, we present the CS-RPMD results with one mapping bead and eight nuclear beads. Under this special limit, CS-RPMD rigorously preserves detailed balance as ${\bf \Gamma}$ becomes an integral of motion of $H_\mathrm{cs}$. However, CS-RPMD is not able to recover the exact QBD even at t=0, due to the fact that there is only one bead for mapping DOF and Eqn.~\ref{eqn:hightz3} becomes a rough approximation. As can be clearly seen, the CS-RPMD correlation functions start to deviate from the exact result even at t=0. Nevertheless, the numerical results of these correlation functions show good agreement to those presented in the main text, suggesting that the nuclear quantization is the most important factor for achieving an accurate result in these test cases. Thus for the calculations presented here, we expect that the MMST Hamiltonian (which appears under this one mapping bead limit) is accurate for describing non-adiabatic effects, and nuclear quantization with ring-polymer should be able to capture the nuclear quantum effects, leading to a reasonable numerical accuracy.
\begin{figure}
 \centering
  \begin{minipage}[t]{1.0\linewidth}
     \centering
     \includegraphics[width=\linewidth]{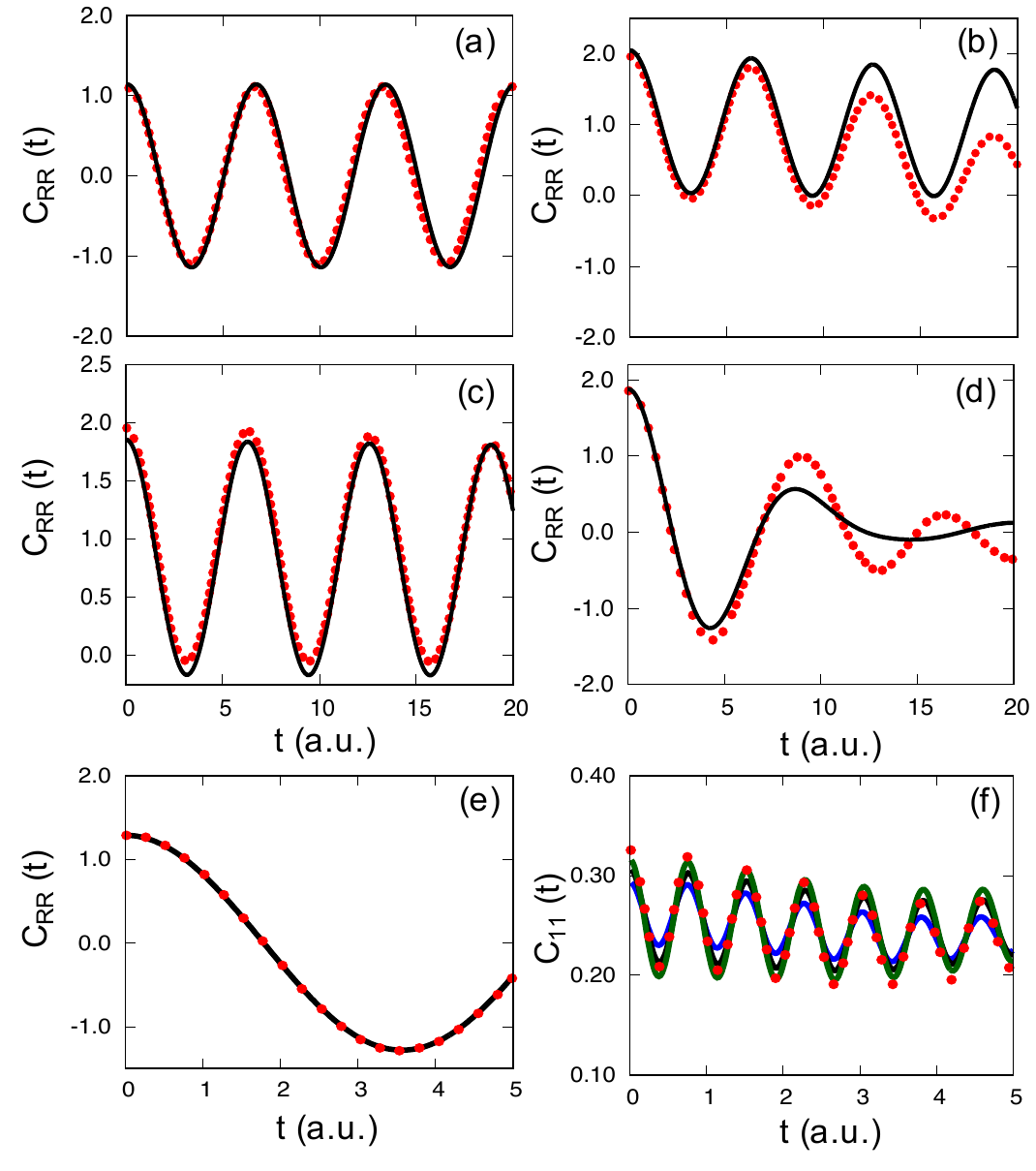}
  \end{minipage}%
   \caption{CS-RPMD correlation functions obtained with one mapping bead and eight nuclear beads. The Kubo-transformed nuclear position auto-correlation functions for model I-IV are presented in (a)-(d), and the Kubo-transformed nuclear position and population auto-correlation functions for model V are presented in (e)-(f). Results are obtained from CS-RPMD (black) and numerical exact DVR method (red). For population auto-correlation function presented in (f), CS-RPMD results are obtained with one mapping bead and 1 (blue line), 8 (black line), and 16 (green line) nuclear beads.}
\label{fig:1bead}
\end{figure}

\bibliographystyle{aipnum4-1}

%




\end{document}